# Digital Advertising Traffic Operation: Machine Learning for Process Discovery

**Massimiliano Dal Mas**
me @ maxdalmas.com

**ABSTRACT**

In a Web Advertising Traffic Operation it's necessary to manage the day-to-day trafficking, pacing and optimization of digital and paid social campaigns. The data analyst on Traffic Operation can not only quickly provide answers but also speaks the language of the Process Manager and visually displays the discovered process problems. In order to solve a growing number of complaints in the customer service process, the weaknesses in the process itself must be identified and communicated to the department. With the help of Process Mining for the CRM data it is possible to identify unwanted loops and delays in the process. With this paper we propose a process discovery based on Machine Learning technique to automatically discover variations and detect at first glance what the problem is, and undertake corrective measures.

## INTRODUCTION

In a Web Advertising Traffic Operation it's necessary to ensure quick and accurate trafficking of campaigns, campaign performance optimizations, troubleshooting, and reporting. Ensure that campaigns launch on time and are fully executed according to the insertion order. [1]

To deliver digital campaigns it's necessary to ensure quick and accurate trafficking of campaigns, campaign performance optimisations, troubleshooting, and reporting and that campaigns launch on time and are fully executed according to the insertion orders. Different processes are involving different activities with advertisers, agencies, and publishers to oversee proper implementation of campaigns (ad tags, beacons, reporting discrepancies, click tracking, etc.) with the retail client as the technical digital media expert.

Moreover it is necessary to maintain proactive communications on account status

---

across multiple account stakeholders including Account Managers, Sales Managers, Senior Management and customer contacts ensuring immediate and direct feedback to both internal and external stakeholders where necessary monitor and improve customer satisfaction as it relates to successful campaign management.

Usually Traffic Managers create document on operations best practices policies and procedures, including an I/O trafficking SLA but trafficking is an ongoing process. Unlike products, processes are less tangible. Processes may only exist in the minds of people and it is difficult to "materialize processes".

In this paper we define a process mining techniques for the trafficking aim to discover, monitor and improve real processes by extracting knowledge from event logs.
We define a technique that automatically discover a process model from event data on a Traffic Office.

## THE BASIC IDEA
Processes can be defined from human behavior partly controlled by procedures and/or information systems. Normative or descriptive models may be used to document processes. But, these models are human-made artifacts that could not necessarily say much about the "real" process performed. As a result, there may be a disconnection between model and reality. Moreover, the people involved in a process are typically unable to understand the corresponding process model and cannot easily see what is actually going on (especially if the products are services or data).

## Process Mining
Process mining combines process models and event data in various novel ways. As a result, one can find out what people and organizations really do. For example, process models can be automatically discovered from event data. Compliance can be checked by confronting models with event data. Loops can be uncovered by replaying timed events on discovered or normative models. Hence, process mining can be used to identify and understand loops, bottlenecks, inefficiencies, deviations, and risks.

## Machine Learning for Process Mining
As part of Artificial Intelligence we can define Machine Learning as the field that "gives computers the ability to learn without being explicitly programmed" [2].
The field of Machine Learning, which aims to develop computer algorithms that improve with experience, holds promise to enable computers to assist humans in the analysis of large, complex data sets [3 – 12]. While Machine Learning collects past information to learn, it is a fact that the past never predicts the future.
It is mirrored in the "Process Mining" approach that assumes that collecting more data on more business processes makes them more predictable. Process Mining is not a reporting tool, but an analysis tool. It enables you to quickly analyze any and very complex processes. Process mining techniques aim to discover, monitor and improve real processes by extracting knowledge from event logs.

## Discovery Algorithm

The process discovery algorithm here presented use a Machine Learning as Process Mining from example behavior recorded in an event log $E$.

This paper is grounded in Linear Temporal Logic (LTL) used to specify constraints on the ordering of activities [13]. We consider an event as composed by different activities that can require different times.

Different activities involved in a Traffic Operation can involve different stakeholders (as: advertisers, agencies, publishers, clients, etc.) having the "working time" as common factor.

For our purpose we define an *interest factor* $f$ as a ratio between the time spent for an activity ($a$) and the time spent for an antecedent activity ($a^{-1}$) and the consequent activity ($a^{+1}$) as show in (1)

$$(1) \quad f = \frac{a}{a^{-1} \times a^{+1}}$$

For an event $E$ a confidence of the constraint is $1$ in a stronger dependency between the antecedent and the consequent activities to reflect the real correlation between the activities [14].

We can use the *interest factor f* to prune processes, e.g., consider only those constraints $f$ above a minimum confidence threshold. A stronger dependency between the antecedent and the consequent is associated with a value of this measure $f$ that is further from $1$ – cause the activity $a$ is waiting the antecedent to be completed [15].

Given an event $E$ consisting of a collection of activities $a$, we could construct a Petri net that "adequately" describes the observed behavior.

We define the *discovery algorithm* as a function $d$ (2) that maps any event $E$ (3) considering a Workflow Net $WN$ (4) and an interest factor $f$ of event $E$ (1).

$$(2) \quad d \in B(u_A) \to u_{WN}$$

$$(3) \quad E \in B(A) \; ; \; A \subseteq u_A$$

$$(4) \quad A_i(WN) = \{a \in \sigma \mid \sigma \in E\}$$

Where $B(A)$ is the set of all the activities and is $u_{WN}$ the set of the activities involved in the Workflow Net $WN$.

Given an *interest factor f* containing events referring to a set of activities $A$ as in (3), we decompose discovery by distributing the activities over multiple sets $A1, A2, \ldots, An$. The same activity may appear in multiple sets as long as $A = A1 \cup A2 \cup \ldots \cup An$.

For each activity set $Ai$, we discover a Workflow Net $WNi$ (4) by applying a discovery algorithm to *interest factor f|Ai*, i.e., the overall *interest factor* projected onto a subset of activities. By combining the resulting Workflow Nets we obtain an overall Workflow Net $WN = WN1 \cup WN2 \cup \ldots \cup WNn$ describing the behavior in the overall event $E$.

## Discovery Loops and delay

It is possible to identify unwanted *loops* in the process defining the set of activities in the Workflow Network $WN$ for multiple sets as long as $A = A1 \cup A2 \cup \ldots \cup An$ as defined below in (5)

$$\text{(5)} \quad loop = \frac{(a_i \cup a_i^-) - (a_i \cup a_i^+)}{a_i^-}$$

## RELATED WORKS

Discovering a process model from a multiset of traces is a very challenging problem and various discovery techniques have been proposed [16 - 27]. Different approaches are used and it is impossible to provide an exhaustive overview of all techniques. The main recent techniques are: heuristics [20, 21], inductive logic programming [22], state-based regions [20, 23], language-based regions [25, 26], and genetic algorithms [27].

## EXPERIMENT

We have built a simulation model reproducing the behavior of event *E*.

For our own interest we did that independent research projects also through specific innovative tasks validate towards average working time declared on "specification required" by the main worldwide industry leading Advertising Agency.

http://www.sizmek-sea.com/Spec/
http://support.adform.com/...specifications/general-specifications/
http://creative-weborama.com/category/uncategorized/
http:// groupm.dk/Banner-sizes-and-specifications

The simulator prototype can describe the evolution of activities over time taking interactions at junctions into consideration. We have investigated the numerical validity of the approximation algorithm. It has proven to be very fast, providing solutions on networks with a few thousands arcs in less than five seconds of CPU time on a PC. We then tested the model's effectiveness using subset of "specification required" by the main worldwide industry leading Advertising Agency.

The fast numerical algorithms described above render the relative optimization problem treatable. This procedure is based on the comparison between measured data and the simulated solutions produced by the algorithm, and gives an average percentage error of about 13%.

## CONCLUSION

In the future, we intend to test the calibration procedure on a more completed workflow network *WN*, and to construct a real-time interface for the automatic representation and visualization of critical activities on the workflow to detect at first glance what the problem is, and undertake corrective measures.

**Massimiliano Dal Mas** is an engineer working on webservices, trafficking and online advertising and is interested in knowledge engineering. In the last years he had to play a critical role at Digital Advertising business, cultivating relationships with key publisher partners with experience managing a team. Been responsible for all day to day operations with partners and consult on the best ways to monetize their properties. His interests include: user interfaces and visualization for information retrieval, automated Web interface evaluation and text analysis, empirical computational linguistics, text data mining, knowledge engineering and artificial intelligence. He received BA, MS degrees in Computer Science Engineering from the Politecnico di Milano, Italy. He won the thirteenth edition 2008 of the CEI Award for the best degree thesis with a dissertation on "Semantic technologies for industrial purposes" (Supervisor Prof. M. Colombetti). In 2012, he received the best paper award at the IEEE Computer Society Conference on Evolving and Adaptive Intelligent System (EAIS 2012) at Carlos III University of Madrid, Madrid, Spain. In 2013, he received the best paper award at the ACM Conference on Web

Intelligence, Mining and Semantics (WIMS 2013) at Universidad Autónoma de Madrid, Madrid, Spain. His paper at W3C Workshop on Publishing using CSS3 & HTML5 has been recommended as position paper.